# Superconducting performance of *ex-situ* SiC-doped $MgB_2$ mono-filamentary tapes

Gennaro Romano[1,*], Arpita Vajpayee[2], Maurizio Vignolo[1], V.P.S. Awana[2] and C. Frederghini[2]

[1] CNR-SPIN, Corso Perrone–24, Genova Italy

[2] Quantum Phenomena and Applications Division, National Physical Laboratory, Dr. K.S. Krishnan Marg, New Delhi-110012, India

Corresponding Author: * Gennaro Romano, CNR-SPIN, Corso Perrone–24, Genova Italy.

Phone no. +39-0106598790; e-mail-gennaro.romano@spin.cnr.it

Key Words: Mono-filamentary *ex-situ* tapes, $MgB_2$, SiC-doping, critical current density.

We report on the superconducting performance of the *ex-situ* SiC doped $MgB_2$ monofilamentary tapes. Polycrystalline powders of $MgB_2$ doped with 5 and 10 wt% SiC were synthesized by conventional solid-state reaction route and characterized for their superconducting performance. It is found that superconducting parameters i.e. upper critical field ($H_{c2}$), irreversibility field ($H_{irr}$) and critical current density ($J_c$) are all improved significantly with SiC addition. Also it was found that relatively lower synthesis temperature (700°C) resulted in further improved superconducting parameters. As synthesized powders are used for *ex-situ* powder-in-tube (PIT) monofilamentary tapes and superconducting parameters are determined. Albeit the superconducting transition temperature ($T_c$) is decreased slightly (~2K) for SiC doped tapes, the superconducting performance in terms of critical current density ($J_c$), being determined from both magnetization and transport measurements, is improved significantly. In particular the SiC doped and 700 °C synthesized $MgB_2$ tapes exhibited the transport $J_c$ of nearly $10^4$ A/cm$^2$ under applied fields of as high as 7 Tesla. Further it is found that the $J_c$ anisotropy decreases significantly for SiC doped tapes. Disorder due to substitution of C at B site being created from broken SiC and the presence of nano SiC respectively in SiC added *ex-situ* $MgB_2$ tapes are responsible for decreased anisotropy and improved $J_c(H)$ performance.

## Introduction

Relatively higher superconducting transition temperature ($T_c$) of up to 40 K, larger coherence length (~10 nm) and its capability to sustain disorder makes $MgB_2$ attractive for superconducting applications [1-4]. In this direction some work is already done in last over nearly decade by now [see ref. 3, 4 and review ref. 5 and 6] since the discovery of superconductivity in $MgB_2$ [1]. Still, though a lot effort is paid for the basic fundamental studies on pure or doped $MgB_2$, the technological impetus of the compound is yet to be realized. For example, the superconducting performance of pristine $MgB_2$ is improved significantly with doping of C or its derivatives [5-8]. In particular the substitution of C at B site in $MgB_2$ brings about the disorder in superconducting hole type sigma band and thus improves significantly the upper critical field ($H_{c2}$) [7,8] and in the case of thin film can reach values as high as 70 Tesla [9]. In this regards, SiC has proven to be one of the most effective additive/substitution to $MgB_2$ for improving its superconducting performance [8-10]. In fact, SiC as an additive not only provides fresh reactive C to substitute at B site and create disorder in superconducting plane but also provide simultaneous pinning centers in form of various nano cluster such as $Mg_2Si$ and nano SiC itself [10,11]. The role of SiC additives in high performance of $MgB_2$ is already explained in various articles [9-11]. It is clear and established by now that SiC as an additive is most effective in improving the superconducting performance of $MgB_2$.

One of the issues to be worked out is to convert the high superconducting performance of $MgB_2$+SiC bulk powders to technological applications in terms of their wires, strands or tapes. In this direction there are several reports in literature [12-15] but yet still no consensus on the issue of how differently the same bulk polycrystalline $MgB_2$+SiC powders compares when drawn in form of tapes or wires. In the present short communication of ours, we intercompare directly our published work on polycrystalline $MgB_2$+SiC sintered powder/pellets [16,17], with that of presently drawn monofilamentary tapes from the same powder. Moreover we compare the results of our ex-situ $MgB_2$+SiC mono-filamentary PIT tapes with the reported similar tapes. We found that though the general effect of improvement in superconducting performance of $MgB_2$ with SiC addition is there in PIT tapes as well, the transformation from bulk to tapes is not complete. The superconducting performance of polycrystalline samples is more profound in comparison to tapes drawn from the same powders. As far as the competitiveness of the studied tapes are concerned our value of critical current as high as $10^4$ A/cm$^2$ under applied fields of as high as 7 T is among the best reported values for ex-situ $MgB_2$ tapes.

## EXPERIMENTAL

Polycrystalline $MgB_2+x$SiC (x = 0, 5wt% & 10wt%, nominally) samples were synthesized by solid-state reaction route. The Mg powder is from *Reidel-de-Haen*, amorphous B powder is from *Fluka* (of assay 95-97%) and SiC nanopowder (n-SiC) is from *Aldrich* (<100 nm). The stoichiometric amounts of ingredients were ground thoroughly, palletized using hydraulic press and put in a tubular furnace for 2.5 hours in argon atmosphere at two different temperatures of 700 °C, and 850 °C. Finally the furnace is cooled in the same atmosphere of argon to room temperature. XRD (X-ray diffraction) is done on *Rigaku miniflex II* powder x-ray diffractrometer using $CuK_\alpha$ radiation. The magnetization/magneto-transport measurements were carried out on *Quantum Design* Physical Property Measurement System (*PPMS*) system. Critical current density ($J_c$) was determined from magnetization loop using the extended Bean model. Fully characterized pallets are subsequently masterized to reduce them in powders for ex-situ powder-in-tube (PIT) process. Monofilamentary $MgB_2$, $MgB_2$+5 wt% SiC and $MgB_2$+10 wt% SiC tapes were fabricated following the *ex-situ* route of the conventional PIT method [18]. $MgB_2$ powders (either pure or doped) were packed inside Ni tubes of 12.7 mm outer and 6mm inner diameter. The tapes were groove rolled and drawn down to a diameter of 2 mm, then cold rolled in several steps to a tape of about 0.35 mm in thickness and 4 mm in width. The superconducting transverse cross section of the conductor was about 0.5 mm$^2$. The conductors were then subjected to a final heat treatment of sintering at 920 °C for 20 minutes in flowing argon atmosphere. Transport critical current density measurements ($J_{ct}$) were performed over 10 cm long samples at the Grenoble High Magnetic Field Laboratory at 4.2 K in magnetic field up to 13 T, applied both perpendicular and parallel to the tape surface, while the current was applied perpendicular to the magnetic field. Short pieces of about 6 mm in length were employed for the magnetization measurements vs. magnetic field performed with a commercial 5.5 MPMS Quantum Design Squid magnetometer. The magnetization measurements were performed at 5 K on part of conductors selected from the same batch of material used for the transport measurements. Demagnetization corrections are negligible at fields above 0.7 T, which is the saturation field of Ni.

## RESULTS AND DISCUSSION

Figures 1(a) and 1(b) show the XRD patterns of 850 °C and 700 °C synthesized bulk polycrystalline $MgB_2$, $MgB_2$+5 wt% SiC and $MgB_2$+10 wt% SiC samples. Small impurity of MgO is seen in all the samples, further with SiC addition the $Mg_2Si$ found along with the presence of some un-reacted raw SiC. The variation of lattice parameters and evolution of various phases including $Mg_2Si$ is discussed in detail in reference 11. It was found that some C (formed with the decomposition of SiC nanopowders) gets substituted at B site and remaining Si reacts with Mg to form $Mg_2Si$ in the $MgB_2$ matrix [11]. Because the same powder will be used subsequently for *ex-situ* PIT tapes, hence it is worth to show the XRD of these powders here once again. Details of structural refinement along with lattice parameters and exact doping of C at B site are discussed in detail elsewhere [11]. Figure 2 represents the resistivity behavior under magnetic field for used pristine $MgB_2$ and $MgB_2$+5 wt% SiC samples being synthesized at 700 °C. It is clear that though the $T_c(R = 0)$ of the SiC doped sample is decreased by about 1.5 K, its performance under magnetic field is improved profoundly. For example $T_c(R = 0)$ of pristine $MgB_2$ under magnetic field of 14 T though decreases to below 7 K, the same is 15 K for the 5 wt% SiC doped sample under same applied field. For quick reference of the powders used, we show here only the behavior of 700 °C synthesized samples and that also for pure $MgB_2$ and 5 wt% SiC doped samples. Other samples data including 10 wt% SiC and various temperatures synthesis are reported earlier [11]. The inter-comparative plots of reduced $T_c$ with magnetic field for the pristine and 5wt% SiC doped and 700 °C synthesized sample are shown in figure 3. It is clear from figures 2 and 3 that with SiC doping the superconducting performance of pristine $MgB_2$ is enhanced profoundly. In fact the calculated upper critical field ($Hc_2$) of the SiC doped samples reaches above 35 T at 0 K [11]. Detailed high resolution transmission HRTEM micro/nano-structural analysis of the SiC doped $MgB_2$ powders is available in reference [11]. Summarily, based upon figures 1 to 3, we can state that the powders used for PIT $MgB_2$ tapes are of reasonably good quality, in particular with SiC addition.

The bulk polycrystalline $MgB_2$ and $MgB_2$+SiC powders described above are drawn in PIT tubes and heat treated in tape form with a final sintering treatment at 920°C for 18minutes. Normalized DC magnetic moment versus temperature plots for these pure and SiC doped $MgB_2$ tapes are shown in figures 4(a) and 4(b). It is clear that superconducting transition temperature ($T_c$) being determined from onset of diamagnetic transition (90% $T_c$) is slightly higher for higher temperature (850 °C) than the lower temperature (700 °C) heat treated powders in the tapes. This is true for both pure and SiC doped samples. Further it is clearly seen that superconducting transition temperature ($T_c$) with SiC doping for both 850 °C and 700 °C heat treated tapes decreases. Also the transition width increases (more than doubled) with increase in SiC content for both synthesis temperatures (850 °C and 700 °C) of the powders used for the tapes. It is clear that some of the added SiC is decomposed and some C is available to substitute B sites and hence a decrease in $T_c$ and an increased transition width is a fingerprint of the disorder introduced in the B- plane. This is in agreement with XRD results on same powders before their use in PIT tapes, as discussed in figures 1(a) and (b), and more details in reference [11].

Critical current density ($J_c$), being extracted from isothermal magnetization (M-H) plots by invoking Beans critical state model for both synthesis temperature (850°C and 700°C) of sintered pure and SiC doped $MgB_2$ tapes at 5K, is plotted in figure 5. The $J_c$ is slightly higher for 700°C synthesized pristine sample below applied magnetic fields of say 4T and, for further higher fields, the same is interestingly lower than the 850°C synthesized sample. Interestingly, the quantity of MgO is nearly the same in both 850°C and 700°C processed pure $MgB_2$ (see figures 1a and 1b). However the crystallinity (grain size) of the two samples is obviously different. The relatively lower temperature 700°C processed tape has small content of un reacted Mg but not present at all in 850°C processed tape. The presence of metallic Mg, possibly at grain boundaries in lower temperature sample, could act like a coupling agent amongst the grains more effectively during the final sintering treatment. Hence, for the tape filled with low synthesis temperature powder this condition corresponds to a better performance of $J_c$ in low fields. In case of SiC doped (5 wt% and 10 wt%) samples being synthesized at 700 °C and 850 °C, no crossover of the $J_c(H)$ is found and moreover the same is improved significantly at least above 3T applied field. The $J_c(H)$ performance is further slightly better for 5 wt% SiC doped and 700 °C processed sample. For a clear view of the impact of the SiC addition on $J_c(H)$ behavior of $MgB_2$, the $J_c(H)$ of variously processed samples at 4T is plotted in the inset of figure 6. It is clear that the $J_c$ of 5 wt% SiC doped sample processed at 700 °C is increased by a factor of three at 4T applied field. In fact the effect is more profound at even higher fields, as the $J_c(H)$ of pristine $MgB_2$ goes down sharply for both synthesis temperature of 700 °C or 850 °C. On the other hand the $J_c(H)$ of SiC doped samples is still linearly decreasing until applied fields of 5 T. In next sections the transport critical current density ($J_{ct}$) will be presented for all the PIT samples under applied fields of up to 12 Tesla. It is clear from the results in figure 6, that though the low field $J_c$ performance is improved only slightly the same is increased by a factor of three under higher applied fields of above 4 T. Higher applied field is required to bring about the resistive vortex motion in SiC doped samples than the pristine ones. This is clear indication that pinning is taking place in SiC doped samples and motion of vortices in applied field is restricted to some extent by the available effective pinning centers. It is discussed in detail in reference [11], that not only some un-reacted SiC, but also inclusions of foreign nano phases such as $Mg_2Si$ are present in SiC doped samples. In fact these several kind of nano phases can act as pinning centers and hinders the motion of vortices in applied fields and thus an effective improvement in $J_c(H)$ of SiC doped samples. As far as the pinning plots are concerned to look for the pinning mechanism, it is really hard to plot the $J_c(H)$ of high performance pinned $MgB_2$ below say 2 T and hence one obviously miss the important pinning peak points. This happens due to flux avalanches taking place in $MgB_2$ [12]. Comparatively we found that $J_c(H)$ being determined from magnetization for our SiC doped PIT tapes is competitive [13-15],. Our (figure 5) $J_c$ for SiC doped $MgB_2$ samples is up to $10^4$ A/cm$^2$ under applied field of 5 T and no abrupt change of behavior at higher field is present. In fact $J_c$ in figure 5 is bulk diamagnetism based critical current density being determined from the magnetization measurements. It will be clear in next sections, when we show the transport $J_c$ in applied fields of above 10 T, that the superconducting performance of SiC doped $MgB_2$ *ex-situ* tapes is improved significantly in comparison to the pristine $MgB_2$ ones till much higher applied fields.

Figure 6 depicts the transport critical current density ($J_{ct}$) against perpendicular applied magnetic field of up to 12 T for all the samples namely $MgB_2$, $MgB_2$+5 wt% SiC and $MgB_2$+10 wt% SiC being synthesized at 700 °C and 850 °C temperatures. As discussed in earlier section for $J_c$ being determined from magnetization measurements, the $J_{ct}$ for low temperature (700 °C) sample is marginally higher than the 850 °C one at lower fields of below 4 T and reverse is the case later for higher fields. Hence we can say the effect of synthesis temperature on transport $J_{ct}$ is the same as for $J_c$ being deduced from magnetization, and is probably related to the grain boundary pinning as explained in the previous section. The $J_{ct}$ for pristine samples comes down to below 2000 A/cm$^2$ at applied fields of about 7 T. It seems the $J_{ct}$ is slightly higher than $J_c$ being determined from magnetization. The reason behind is due not only to the very different values of the electric fields applied in magnetization and transport measurements but also the strong dependence of the same on the specific microstructure of $MgB_2$ [16]. In any case the $J_{ct}$ of pure $MgB_2$ mimics the magnetization $J_c$ measurements on the same samples. With SiC doping the in-plane $J_{ct}$ is improved significantly and is more than 8000 A/cm$^2$ at around the same field of 7 T. In fact the $J_{ct}$, which is dying at above 7 to 8 T for pure $MgB_2$ is significant up to 10-12 T applied fields for SiC doped tapes. Quantitatively the inset of figure 6 shows the transport critical current at 7 T applied field for variously processed (700 °C and 850 °C) samples. It is found that for $MgB_2$+5 wt% SiC sample synthesized at 700 °C the $J_{ct}$ is increased of a factor of four when compared with the undoped $MgB_2$ of the same batch. Hence it is clear from results being exhibited in figure 7, that the transport critical current density ($J_{ct}$) is significantly improved for SiC added *ex-situ* $MgB_2$ PIT monofilamentary tapes. It is obvious from figures 5 and 6 that we have been able to transfer the high superconducting performance effect of SiC doping from the powders (figures 1-4) to the *ex-situ* mono-filamentary tapes being drawn from the same powders and post heat treated (sintering) for a good connectivity. The critical transport current density ($J_{ct}$) being reported earlier for pinned (SiC/C added) $MgB_2$ wires, strands or tapes [13-15] are competitive with our value of 8000 A/cm$^2$ at around 7 T field. On the other hand we still feel that there is further scope to improve upon the presently obtained values, in particular by optimizing the $MgB_2$ synthesis and the tape processing parameters.

One of the important issues on the $MgB_2$ tapes is related to the anisotropic behavior of the critical current. This is basically due to the intrinsic anisotropy of the $MgB_2$ itself, arising from its two band (sigma and pi) electronic structure. Though the Fermi surface is mostly (90%) occupied by superconducting sigma band the role of minor (10%) electron type normal pi band cannot be ignored [17, 18]. The $J_c$ transport measurements (both in plane and out of plane) reveal the effect of SiC doping on the anisotropy. The less is anisotropy and the better and smooth is the current flow. One of the effective way to decrease the anisotropy of $MgB_2$ is to disorder its dominant sigma band, for which B site C substitution is the ideal choice [9,20]. In fact sigma band is formed from B plane. The current strategy of SiC addition to $MgB_2$ is the obvious choice, because of the release of some fresh C during synthesis and thus effective B site C substitution in host $MgB_2$. To check the hypothesis, we present in figure 8 and 9 the anisotropy measurements on our various temperature synthesized $MgB_2$, $MgB_2$+5 wt% SiC and $MgB_2$+10 wt% SiC *ex-situ* tapes.

Figure 7 represents the in (//ab) and out (//c) of plane $J_{ct}$ transport measurements plots against magnetic field for $MgB_2$, and $MgB_2$+10 wt% SiC tapes being synthesized at 700 °C. As seen from this figure the $J_{ct}$ is higher for in plane in comparison to out of plane measurements. This indicates that during the rolling process some alignment of the polycrystalline powder takes place in PIT tapes. Also it is clearly seen that $J_{ct}$ is significantly higher in both in plane and out of plane measurements for SiC doped samples, similar to that as observed in figures 5 and 6. To further elucidate on the effect of anisotropy, in the inset of figure 7, we present the anisotropy versus magnetic field plot for both pure and SiC doped samples. It is clear from this inset that anisotropy is reduced significantly by nearly an order of magnitude at 7T for the SiC doped PIT tapes. The anisotropy measurements for $MgB_2$ and $MgB_2$+5 wt% SiC tapes being synthesized at 850 °C PIT mono-filamentary tapes are presented in figure 8. The effect of the anisotropy is more or less qualitatively the same as for the 700 °C synthesized PIT tapes being shown in figure 7. It is clear that the SiC addition did not only improved the $J_{ct}$ significantly but also reduce the anisotropy by an order of magnitude for both 700 °C and 850 °C synthesized SiC doped PIT mono-filamentary tapes. This is important that one not only desires to get the higher $J_{ct}$ but the reduced anisotropy as well. The origin of this less anisotropy could be the same as discussed in previous paragraph, that is the substitution of C at B site being taking place in SiC doped PIT tapes. This effect has its origin related to complex electronic band structure of $MgB_2$ [17,18].

Conclusions

In summary, we synthesized high superconducting bulk polycrystalline $MgB_2$ added with SiC and turned it into powder form for *ex-situ* PIT mono-filamentary tapes. It is found that both $J_c$ (magnetic) and $J_{ct}$ (transport) are improved significantly in particular under magnetic fields to the tune of around 8000 A/cm$^2$ at 7T. Further the anisotropy is reduced by an order of magnitude for SiC added PIT $MgB_2$ *ex-situ* mono-filamentary tapes. The origin of the profound improvement of the superconducting performance of SiC added PIT $MgB_2$ *ex-situ* mono-filamentary tapes is explained on the basis of C substitution on B sites reached in pristine $MgB_2$, with a contemporaneous presence of pinning centers; i.e., nano SiC and $Mg_2Si$ present in the matrix as shown by XRD patterns. It is further to be noted that our SiC added PIT $MgB_2$ *ex-situ* mono-filamentary tapes possess competitive or even better $J_c(H)$ and $J_{ct}(H)$ performance than the ones reported in literature.

FIGURE CAPTIONS

Figure 1. XRD (X-ray diffractions) for 850 $^0$C and 700 $^0$C processed various pure $MgB_2$, $MgB_2$+5wt% SiC and $MgB_2$+10wt% SiC samples.

Figure 2. Resistivity under magnetic field ρ(T)H plots for pure and 5wt%SiC doped $MgB_2$ samples synthesized at 700 $^0$C.

Figure 3. Reduced superconducting transition temperature $T_c$(R=0)H divided by $T_c$(R=0)at zero field plots for pure and 5wt%SiC doped $MgB_2$ samples synthesized at 700 $^0$C.

Figure 4. Normalized moment (m) versus temperature plots in superconducting regime (below 42K) for (a) 700 $^0$C and (b) 850 $^0$C processed $MgB_2$, $MgB_2$+5wt%SiC, and $MgB_2$+10wt%SiC samples.

Figure 5. Critical current density ($J_c$) determined from magnetization measurements versus applied field plots for 700 $^0$C and 850 $^0$C processed $MgB_2$, $MgB_2$+5wt%SiC, and $MgB_2$+10wt%SiC samples at 5K. Inset shows the $J_c$ at 4T versus added SiC concentration plot for the same samples, with three time increase in $J_c$ (4T) for 5wt%SiC doped samples at 5K.

Figure 6. In plane (ab) transport critical current density ($J_{ct}$) versus applied field plots for 700 $^0$C and 850 $^0$C processed $MgB_2$, $MgB_2$+5wt%SiC, and $MgB_2$+10wt%SiC PIT monofilmentary *ex-situ* tapes at 5K. Inset shows the $J_{ct}$ at 7T versus added SiC concentration plot for the same samples, depicting five time increase in $J_{ct}$ (7T) for 5wt%SiC doped samples at 5K.

Figure 7. Both in (//ab) and out (//c) of plane transport critical current density ($J_{ct}$) versus applied field plots for 700 $^0$C processed $MgB_2$, and $MgB_2$+10wt%SiC PIT monofilmentary ex-situ tapes at 5K. The inset shows the magnetic field dependence of the anisotropy for the same tapes.

Figure 8. Both in (//ab) and out (//c) of plane transport critical current density ($J_{ct}$) versus applied field plots for 850 $^0$C processed $MgB_2$, and $MgB_2$+5wt%SiC PIT monofilmantary *ex-situ* tapes at 5K. The inset shows the magnetic field dependence of the anisotropy for the same tapes.

Figure 1

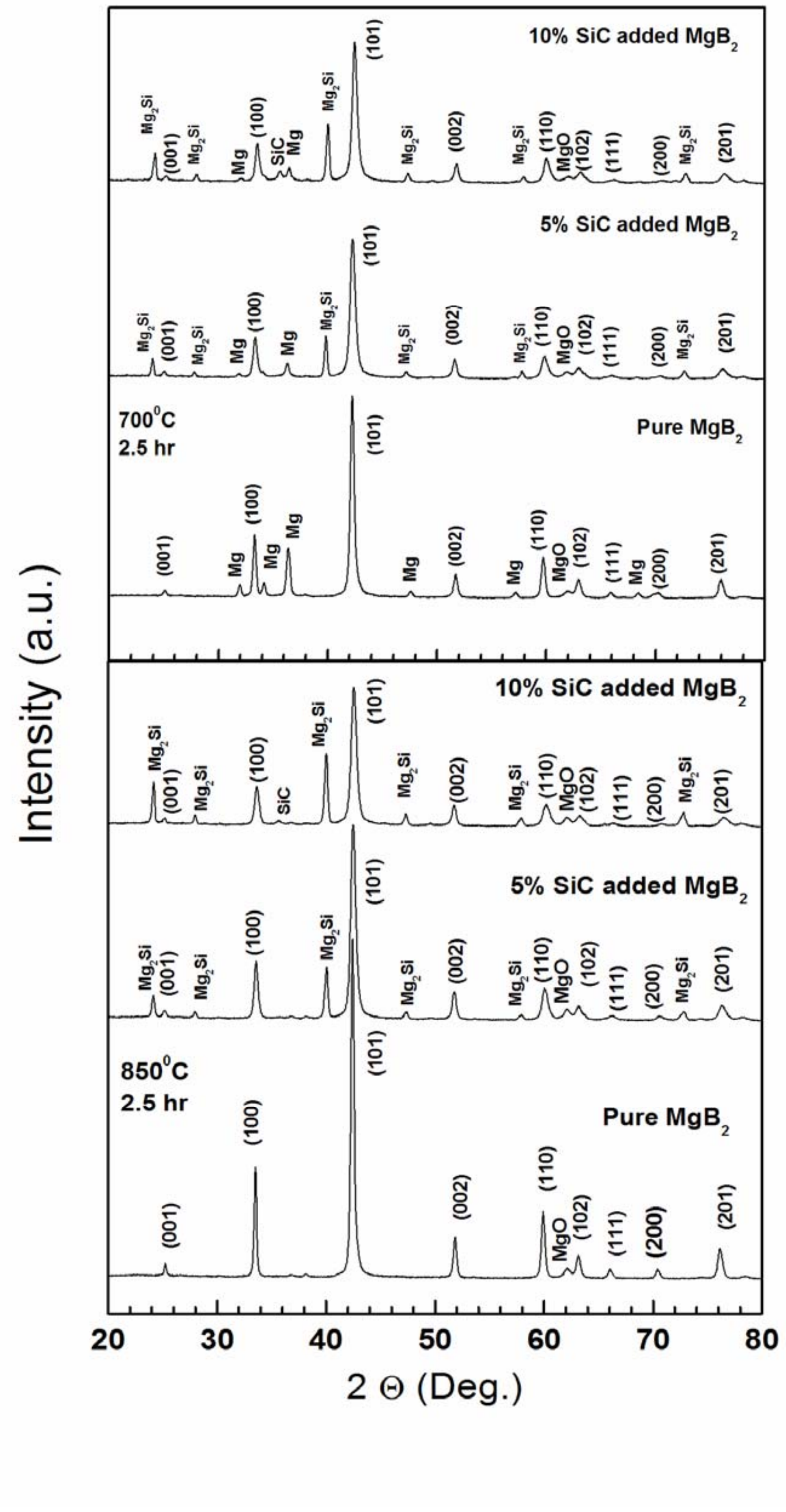

Figure 2

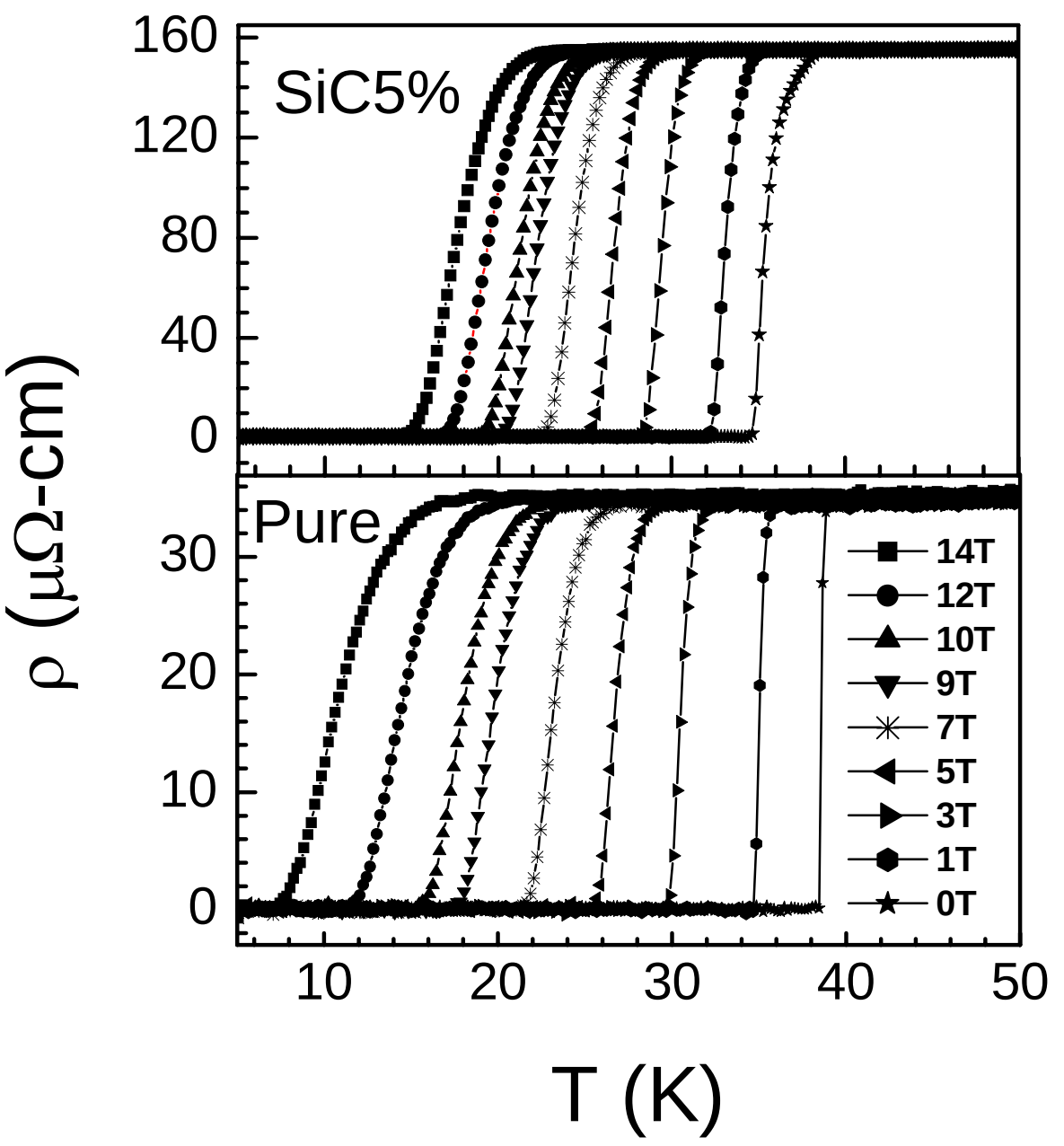

Figure 3

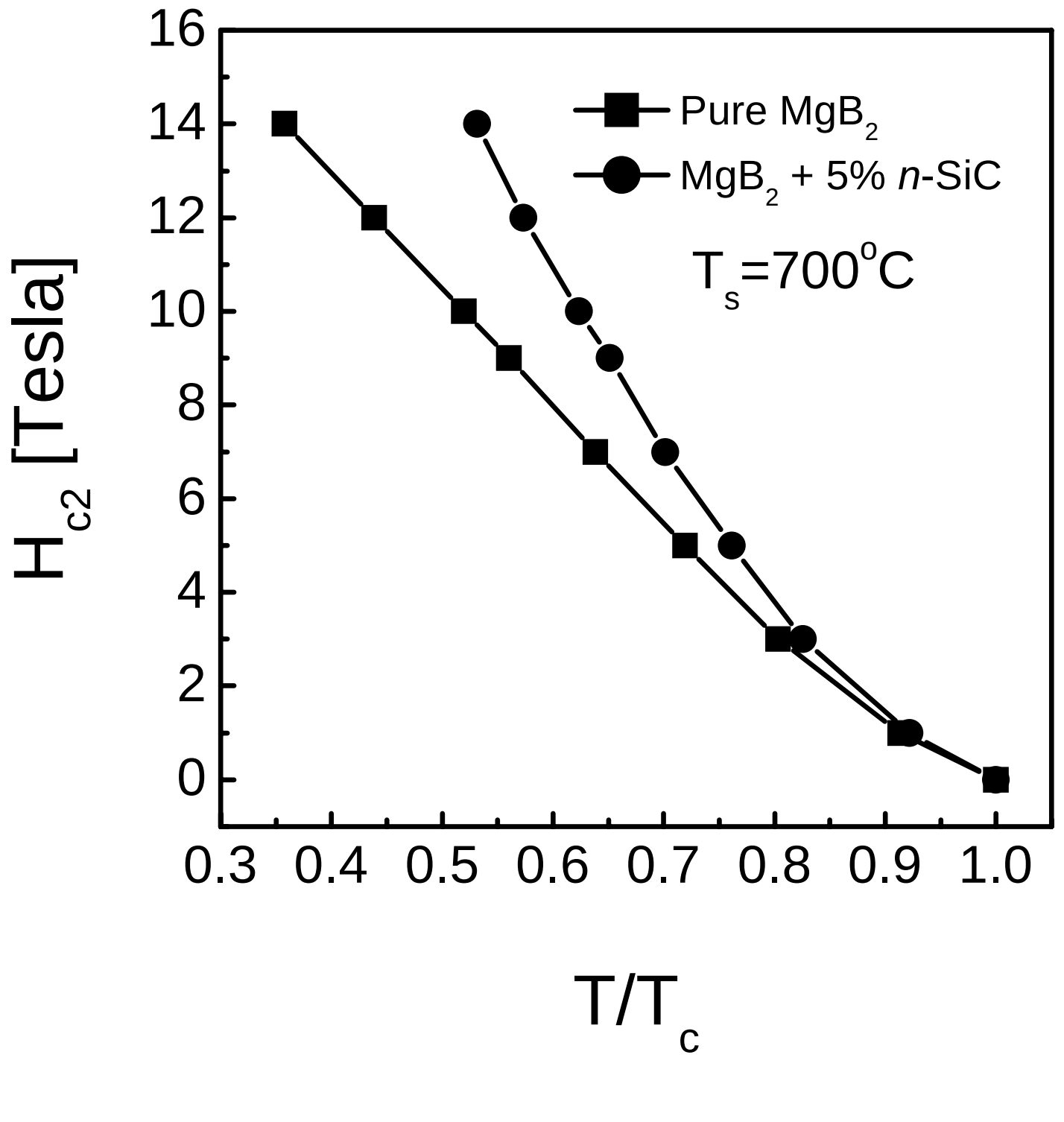

Figure 4.a

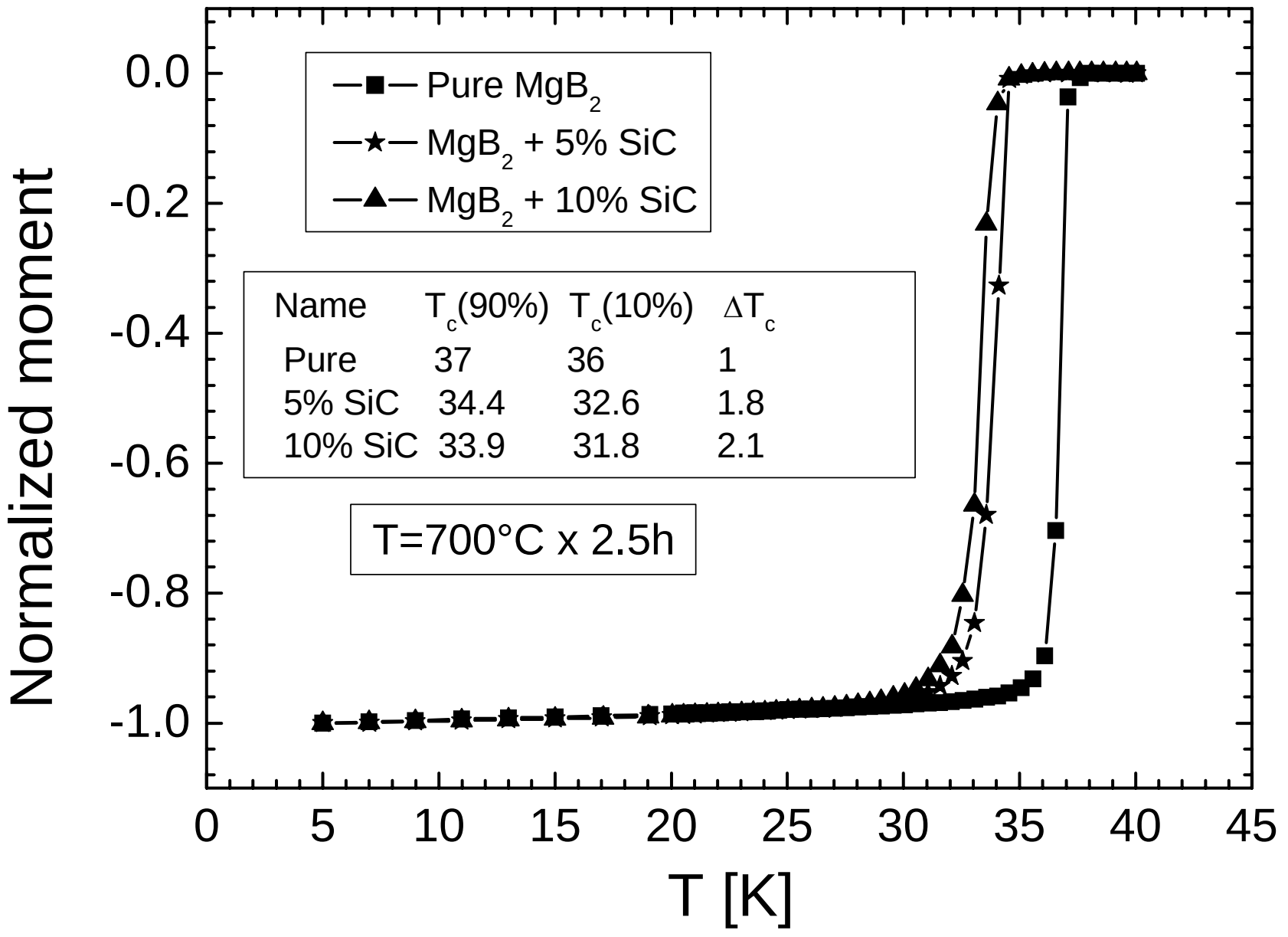

Figure 4.b

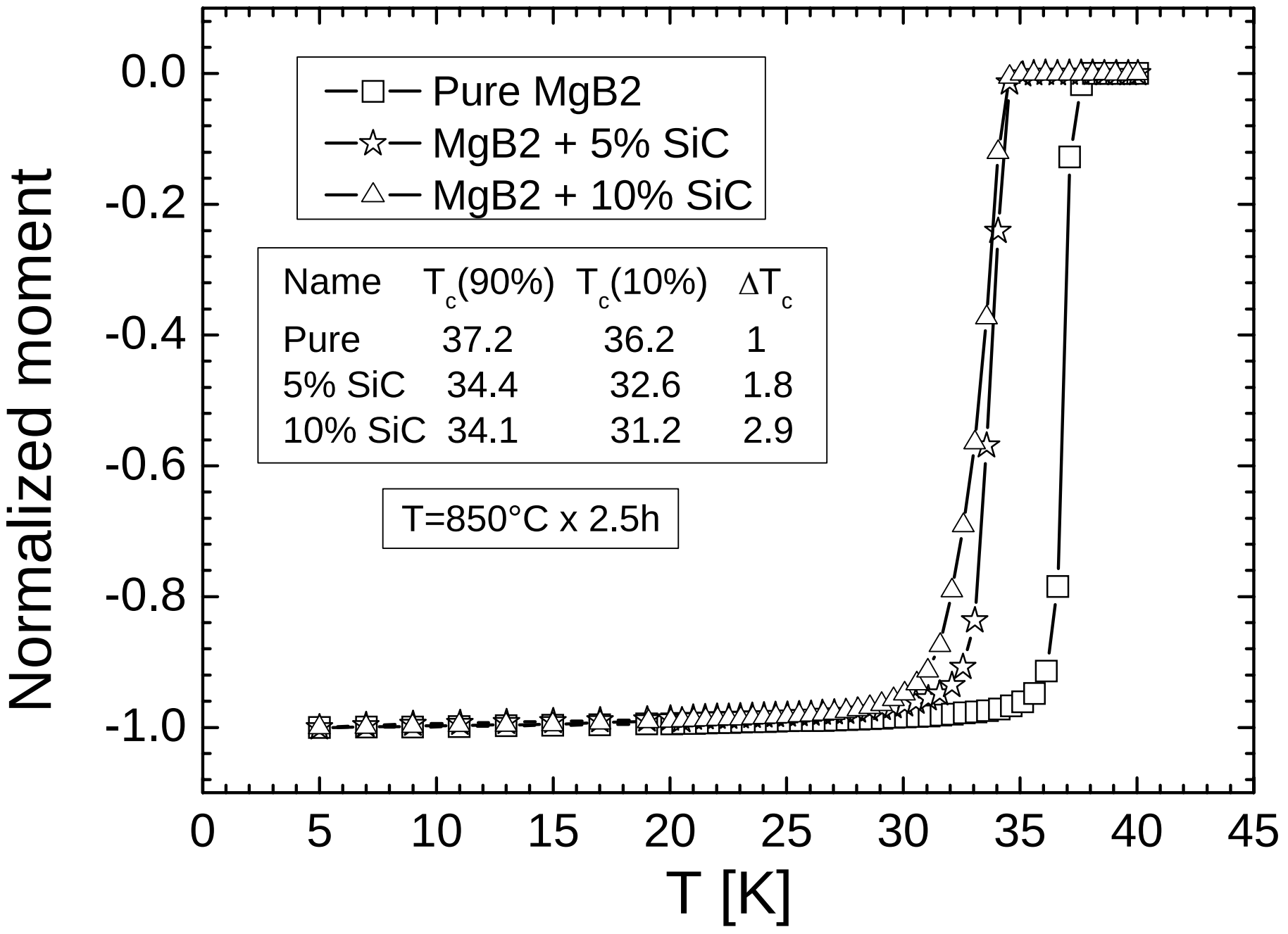

Figure 5

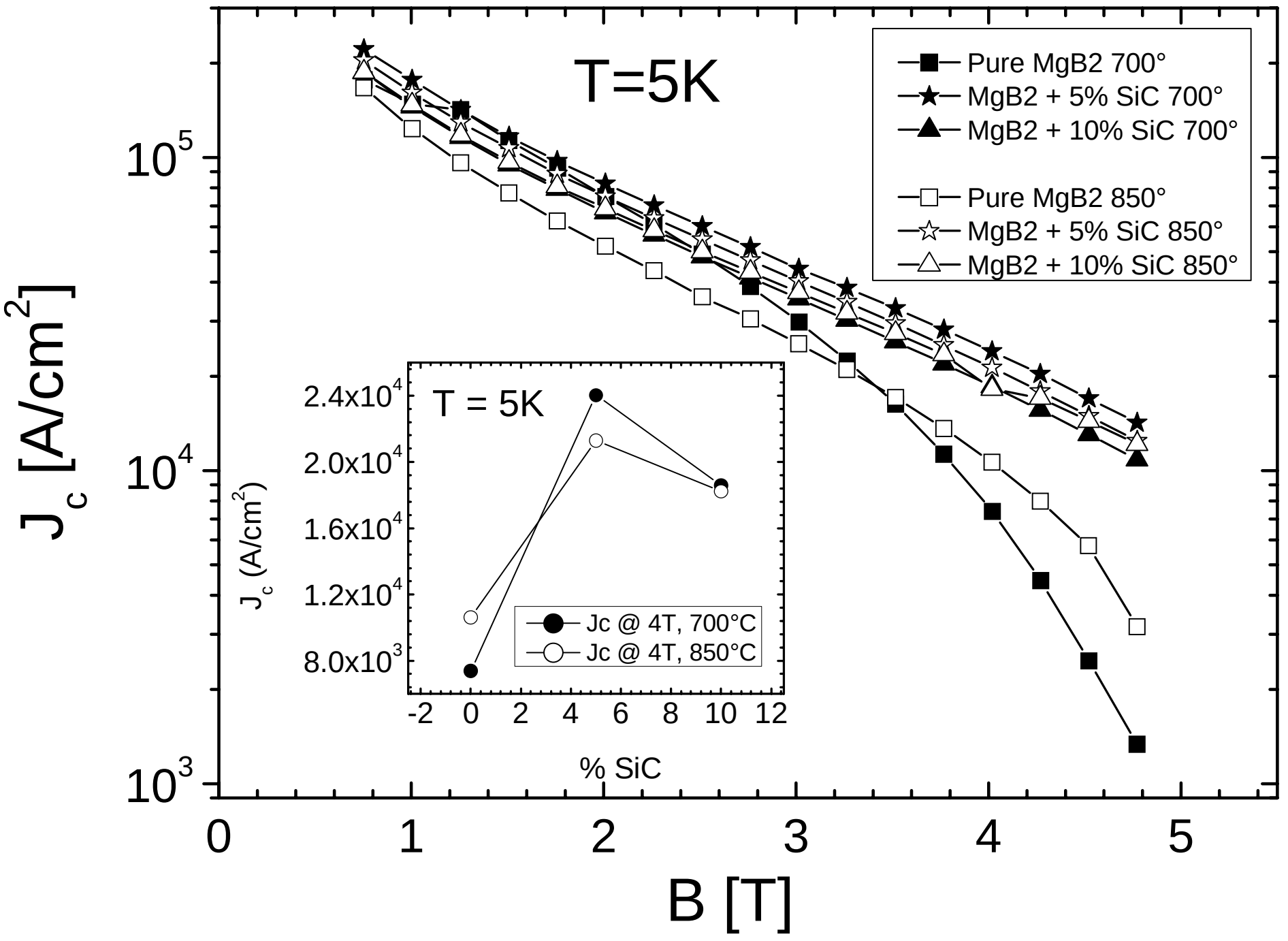

Figure 6

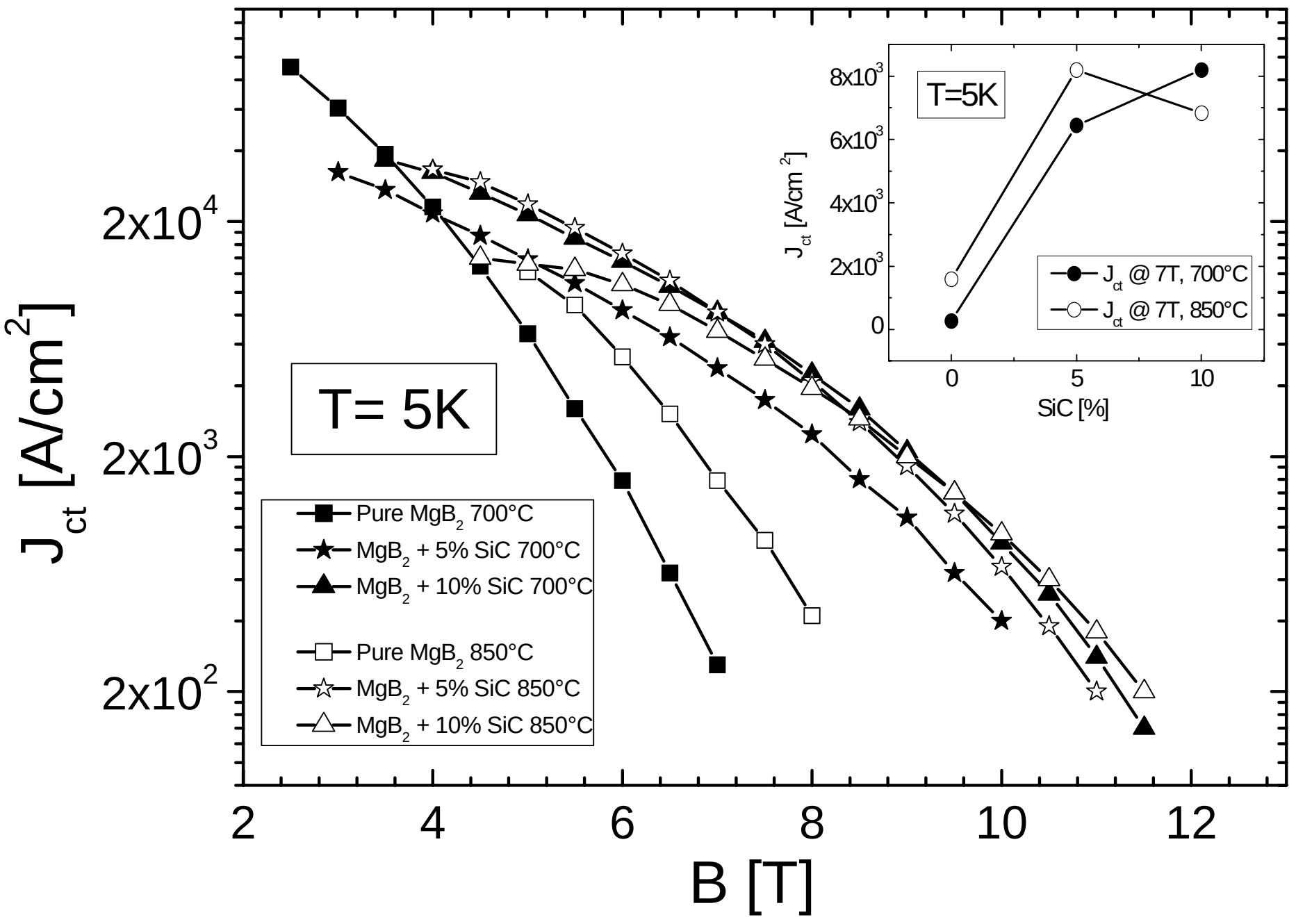

Figure. 7

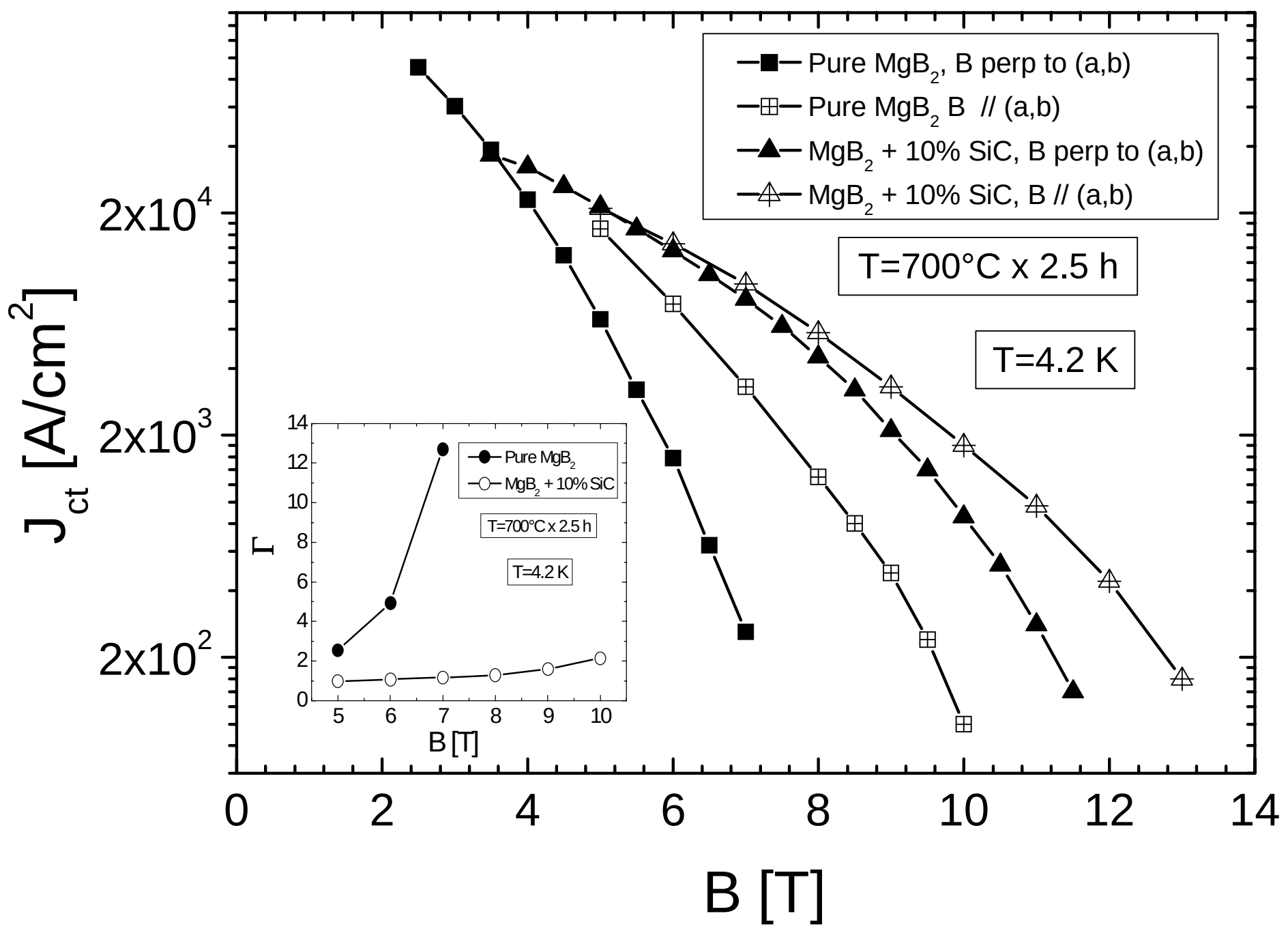

Figure.8

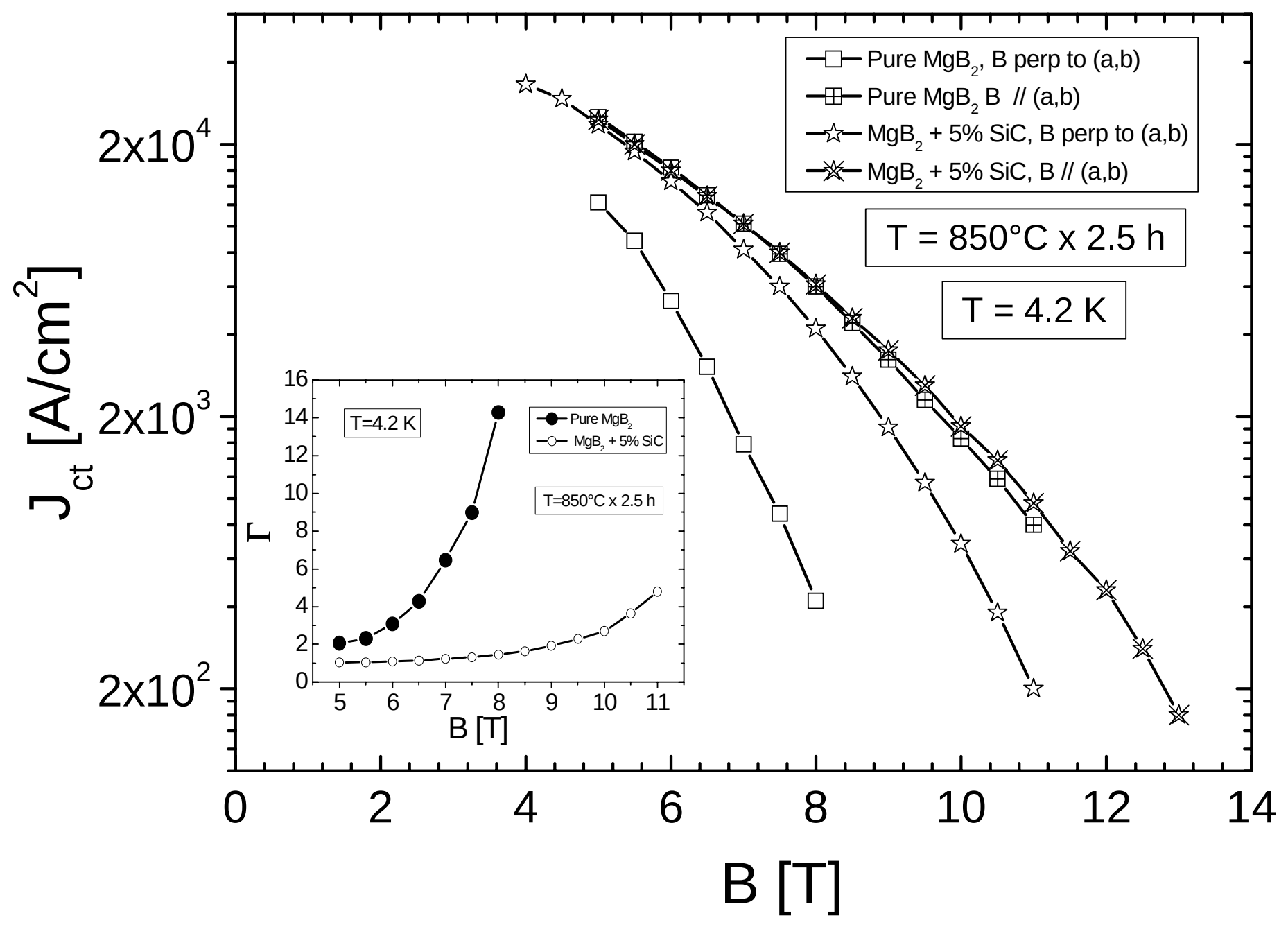